\documentclass[fleqn,twoside]{article}
\usepackage{espcrc2}
\usepackage{graphicx}
\usepackage{epsfig}
\usepackage[figuresright]{rotating}

\newcommand{\AmS}{{\protect\the\textfont2
  A\kern-.1667em\lower.5ex\hbox{M}\kern-.125emS}}
\title{Prospects for Higgs search at the LHC}

\author{F. Piccinini\address[CERN]{CERN, Theoretical Physics Division, 
                                   CH~1211 Geneva 23, Switzerland}%
        \thanks{On leave  of absence from INFN Sezione di Pavia, Italy.}
        \thanks{Work supported in part by the EU Fourth 
                Framework Programme ``Training and Mobility of Researchers'', 
                Network ``Quantum Chromodynamics and the Deep Structure of 
                Elementary Particles'', contract FMRX--CT98--0194 
                (DG 12 -- MIHT).}}
       
\begin{document}

\begin{abstract}
The present status of theoretical calculations for signal and background 
processes relevant to the Higgs boson search at the LHC is reviewed, with 
special emphasis on recent developments. 
The issue of Higgs properties determination at the LHC is addressed.
\vspace{1pc}
\end{abstract}

\maketitle

\leftline{{\tt Preprint numbers: CERN-TH/2002-266,}}
\leftline{{\tt \hskip 3.03truecm \, \, FNT/T 2002/15}}

\section{Introduction}
The search for the Higgs boson will be one of the primary goals of the 
LHC experiments.  
The main Higgs production channels at the LHC are gluon fusion, weak boson 
fusion (WBF) and $t \bar t H$ and $V H (V=W,Z)$
associated productions. Complete analyses with full detector simulation 
within the experimental collaborations have shown that, 
with only 10~fb$^{-1}$ of integrated luminosity per experiment, 
corresponding to one year of running at low luminosity 
(${\cal L} = 10^{33}$ cm$^{-2}$ s$^{-1}$), there is a signal significance 
above the $5 \sigma$ level over a wide range of Higgs masses, the most difficult 
region being the window between the lower bound given by LEP and 
140~GeV~\cite{atl99-15}. 
Recent detailed simulations, with realistic experimental conditions of the WBF 
processes, have proved the importance of these channels to improve significantly 
the sensitivity in that difficult region~\cite{jakobs,zeyrek}. 
Crucial to this aim, on the experimental 
side, are the forward jet reconstruction and central jet veto efficiencies. 
On the theoretical side, a large amount of work has been done to reduce the 
uncertainties on theoretical predictions, both for the signals and the 
backgrounds. 
Further studies have been performed to improve on the strategy originally 
proposed in ref.~\cite{zknrw00} for the determination of Higgs boson properties, 
such as couplings to fermions and gauge bosons, and total width. Moreover, very 
recently, the first analyses on the LHC potential for the Higgs self-coupling 
determination have been carried out. In the following the present status of 
theoretical calculations and of the prospects for Higgs properties determination 
is reviewed. The main focus will be on the mass window 115-200~GeV, which 
is the preferred one by electroweak precision data and also 
partially by supersymmetry. 
During the last year a lot of effort has been concentrated on 
the Standard Model 
(SM) Higgs boson, but the strategies and results can be translated to the 
lightest scalar Susy Higgs. For these reasons this brief review deals only with 
the case of the SM Higgs boson.

\section{Theoretical calculations}
In order to disentangle a signal from backgrounds, 
a good understanding of uncertainties in theoretical predictions is necessary. 
Predictions based on leading order (LO) calculations 
are plagued by considerable uncertainties due to the strong dependence on 
the renormalization and factorization scales, introduced by the QCD coupling 
and the parton densities. 
At present the QCD corrections, at least at next-to-leading order (NLO), are 
known for all production channels, the most recent calculations being the 
NNLO calculation for the gluon fusion process in the limit of heavy top-quark
mass~\cite{grazzini,kilgore} and the NLO corrections for the process 
$pp / p \bar p \to t \bar t H + X$~\cite{tthnlo}. The higher-order corrections 
reduce the renormalization and factorization scale dependence of the 
theoretical predictions. 
In the case of $pp / p \bar p \to t \bar t H + X$, the NLO corrections increase 
the rate by roughly 20\% with respect to the LO predictions over the entire 
intermediate Higgs mass range at the LHC. In the case of Higgs production 
through gluon fusion, the NLO corrections give a $K$ factor of the order of 2, 
so that NNLO corrections are needed~\cite{grazzini,kilgore}. 
At present, the uncertainties arising from QCD uncertainties (combining the 
residual scale dependence with the error from parton distribution functions) 
can be estimated to be of the order of $\pm$20\% for gluon fusion, $\pm$5\% for 
WBF, $\pm$10\% for associated production. 

Concerning the backgrounds, several processes with low final state parton
multiplicity (corresponding to important irreducible backgrounds) are available 
at NLO, namely 
$q \bar q \to \gamma \gamma$~\cite{diphox}, 
$g g \to \gamma \gamma$~\cite{gluglugg}, 
$p p(\bar p) \to W b \bar b$, $p p(\bar p) \to Z b \bar b$~\cite{evc}, 
$p p(\bar p) \to W j j$, $p p(\bar p) \to Z j j$~\cite{ce}, 
$p p(\bar p) \to VV$~\cite{fw} and 
QCD $H + jj$ production via gluon fusion~\cite{qcdhjj}. Some of these 
calculations are already implemented in NLO Monte Carlo programs. 
In the case of multiparton final states, the methods developed up to now 
for NLO calculations cannot be applied, because of the complexity of the 
calculations for processes with many external legs. Recently some 
effort 
has been devoted to the realization of LO Monte Carlo event generators 
based on exact matrix element calculations, suitably interfaced to the shower 
evolution Monte Carlo programs producing the real final state 
hadrons~\cite{alpgen,comphep,grace,madgraph,acermc}.

\section{Higgs couplings to fermions and gauge bosons}
The LHC will allow not only the discovery of the Higgs boson, but also the 
study of its properties, such as mass, width and couplings to 
fermions and gauge bosons. While the decay channels $H \to \gamma \gamma$ 
and $H \to Z Z^{(*)} \to 4l$ will allow a direct mass measurement at 
the 0.1\% level over a wide range of masses, the total width can only 
be determined with about 10\% accuracy by direct measurement with the decay 
$H \to Z Z^{(*)} \to 4l$ for $m_H > 200$~GeV, the Higgs width 
for lower Higgs masses being too small with respect to the detector resolution. 
As will be shown below, by exploiting the available production and decay 
mechanisms at the LHC, an indirect measurement of the total 
width can be performed also in the low mass region. 
\begin{figure}[htb]
\includegraphics[width=0.46\textwidth,clip]{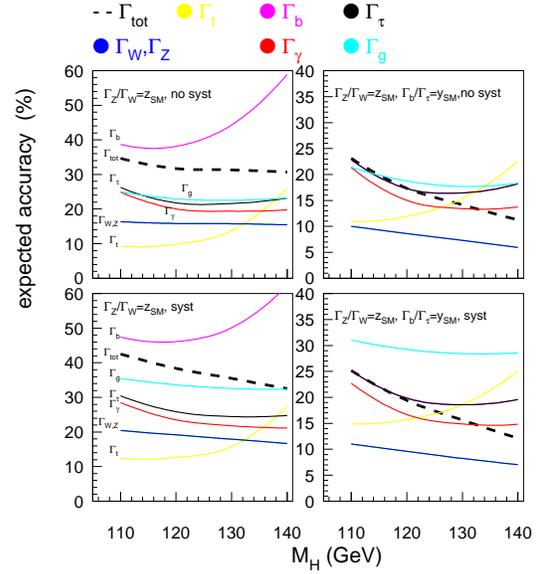}
\vspace{-35pt}
\caption{Relative accuracy (\%) on the individual rates $\Gamma_i$ expected 
at the LHC (from ref.~\cite{rb02}). See the text for a detailed description 
of the panels.}
\label{fig:fig1}
\end{figure}
In principle, the Higgs coupling, for instance to a given fermion family $f$, 
could be obtained from the following relation:
\begin{eqnarray}
R(H\to f \bar f)&=&\int{L dt}\cdot \sigma(pp\to H)\cdot 
{\frac{\Gamma_f}{\Gamma}}, \nonumber
\end{eqnarray}
where $R(H\to f \bar f)$ is the Higgs production rate in a given final state, 
which can be measured experimentally, $\int{L dt}$ is the integrated 
luminosity, $\sigma(pp\to H)$ is the Higgs production cross section, 
and $\Gamma$ and $\Gamma_f$ are the total and partial Higgs widths 
respectively. Hence, a measurement of the Higgs 
production rate in a given channel allows the extraction of the partial
width for that channel, and therefore of the Higgs coupling $g_{f}$ to
the involved decay particles ($\Gamma_f\sim g_f^2$), provided that the
Higgs production cross-section and the total Higgs width are known
from theory. 
Aiming at model-independent coupling determinations, one needs to 
consider ratios of couplings, which are experimentally accessible
through the measurements of ratios of rates for different final
states, because in the ratio the total Higgs cross-section and width
cancel (as well as luminosity and all QCD uncertainties related to the initial 
state). In spite of the fact that the gluon fusion mechanism is 
the leading scalar Higgs production mode at the LHC, other subleading 
production modes, such as weak boson fusion and associated production, are 
extremely important to provide complementary information and allow 
unique determinations of ratios of Higgs boson couplings. 
Up to now detailed studies on signal and backgrounds for several channels 
have been performed, namely 
$gg \to H, (H \to \gamma \gamma, ZZ, 
WW)$~\cite{atl99-15,cms94-38,dreiner,denegri}, 
$qq \to qqH, (H \to \gamma \gamma, \tau \tau, 
WW)$~\cite{rz97,rzh99,prz00,rz99,kprz01}, 
$p p \to t \bar t H, (H \to b \bar b, WW, 
\tau \tau)$~\cite{rws99,dmd01,mrw02,rb02} and 
$p p \to W H, H \to b \bar b$~\cite{dmd02}. Each process depends on two 
Higgs couplings, one from the Higgs boson production and one from the Higgs 
boson decay, with the exception of the weak boson fusion channels, 
for which it is experimentally 
impossible to distinguish between $WW \to H$ and $ZZ \to H$ production 
mechanisms. However, 
since the couplings of a scalar Higgs boson to the $Z$ and $W$ gauge bosons are
closely related to the electroweak $SU(2)$ gauge symmetry, which has been
very successfully tested by the LEP experiments, and since in a large class 
of models the ratio of $HWW$ and $HZZ$ couplings is identical to the one in the 
SM, including the MSSM, it is reasonable to assume 
$\Gamma_Z / \Gamma_W = z_{SM}$. Under this hypothesis, every production and
decay channel provides a measurement of the ratio $Z_j^{(i)} = 
\Gamma_i \Gamma_j / \Gamma$, where $i = g,W,t$ indicates the production process 
and $j=b,\tau,W,Z,g,\gamma$ indicates the decay process. For the case 
$m_H < 140$~GeV, the above mentioned channels allow us to express the individual
rates $\Gamma_t$, $\Gamma_b$, $\Gamma_\tau$, $\Gamma_W$, $\Gamma_g$ and
$\Gamma_\gamma$ as functions of the observables $Z_j^{(i)}$ and the total 
Higgs width $\Gamma$~\cite{rb02}. With the additional assumption that the total 
width is saturated by the known channels $\Gamma = \Gamma_b + \Gamma_\tau 
+ \Gamma_W + \Gamma_Z + \Gamma_g + \Gamma_\gamma$ (otherwise new processes 
would be observed independently of any precision study), an expression for 
$\Gamma$ can be obtained in terms of measured quantities, namely~\cite{rb02}
\begin{eqnarray}
\sqrt{\Gamma} &=& \frac{1}{\sqrt{Z_W^{(W)}}} \left[ Z_\tau^{(W)} 
\left( 1 + \frac{Z_b^{(t)}}{Z_\tau^{(t)}}\right) \right. \nonumber \\ 
         &+& \left. Z_W^{(W)} (1 + z_{SM}) 
         + \frac{Z_W^{(W)} Z_\gamma^{(g)}}{Z_\gamma^{(W)}} + Z_\gamma^{(W)}
\right] . \nonumber
\end{eqnarray}
Figure~\ref{fig:fig1}~\cite{rb02} summarizes the relative accuracy on the 
individual rates $\Gamma_i$ expected in the model-independent scenario as 
well as in a scenario with $\Gamma_b/\Gamma_\tau$ fixed to its SM value, 
assuming a total integrated luminosity of 200~fb$^{-1}$. 
The upper plots show the accuracies obtained without including any 
theoretical systematic error, while the lower plots show the same accuracies 
when a systematic theoretical error of $20\%$ for the $gg\to H$ channel, of
$5\%$ for the $qq\to qqH$, and of $10\%$ for the $pp\to t\bar{t}H$
channel are included. As can be seen, the total Higgs width can be indirectly 
determined in the low mass region with a precision of the order of 30\% in a 
model-independent way, and the Higgs couplings can be determined with 
accuracies between 7\% and 25\%. In the case of $140 < m_H < 200$~GeV, the 
gluon fusion, weak boson fusion and $t \bar t H$ associated production 
processes can be exploited, with the Higgs 
boson decaying only to gauge bosons, allowing an indirect determination of 
$\Gamma_W$ and $\Gamma$ with a precision of the order of 10\%~\cite{zknrw00,z02}. 
In this Higgs mass range, however, there is no handle to study the Higgs Yukawa 
coulings to $b$ quarks and $\tau$ leptons. 
The assumption $\Gamma_Z / \Gamma_W = z_{SM}$ 
can be tested at the 20--30\% level for $m_H > 130$~GeV by measuring the ratio 
$Z_Z^{(g)} / Z_W^{(g)}$~\cite{z02}, and it can even be tested with the 
same level of accuracy for lower 
Higgs boson masses by comparing the two ratios $Z_b^{(WH)} / Z_b^{(t)}$ and 
$Z_\tau^{(W)} / Z_\tau^{(t)}$~\cite{rb02}. For $m_H > 140$~GeV, 
with luminosities of the order of 
300~fb$^{-1}$, the ratio $\Gamma_t/\Gamma_g$ can be tested in a 
model-independent way through a measurement of 
$Z_W^{(t)}/Z_W^{(g)}$~\cite{mrw02}. 

\section{Higgs self-couplings}
A complete determination of the parameters of the SM would 
require the measurement of the Higgs self-couplings. These include trilinear 
and quadrilinear interactions. In the SM the corresponding 
couplings are fixed at LO in terms of the Higgs mass and vacuum expectation
value $v$, namely $\lambda_{HHH}^{SM} = 3 m_H^2 / v$, 
$\lambda_{HHHH}^{SM} = 3 m_H^2 / v^2$. A direct measurement of $\lambda_{HHH}$ 
could be obtained via the detection of Higgs pair production, where a 
contribution is expected from the production of a single off-shell Higgs which
decays into a pair of Higgses. This contribution is always accompanied by 
diagrams where the two Higgs bosons are radiated independently, with couplings 
proportional to the Yukawa couplings or the gauge couplings. As a result, 
different production mechanisms will lead to different sensitivities of 
the $HH$ rate to the value of $\lambda_{HHH}$. In the literature the following 
SM channels have been considered~\cite{dkmz99}: 
inclusive $HH$ production, dominated by the 
partonic process $gg \to HH$; 
vector boson fusion $qq \to qqV^*V^* \to qqHH$, 
associated production with $W$ or $Z$ bosons $q \bar q \to V H H$; 
associated production with top-quark pairs $gg/q\bar q \to t \bar t HH$. 
With the exception of the gluon fusion process, which has a total cross section 
at the level of few tens of fb, the cross section for all other channels is 
of the order of 1~fb over the intermediate Higgs mass range~\cite{dkmz99}. 
Given these low
production rates, and the potentially large backgrounds associated to the $HH$ 
final states, a quantitative study of the Higgs self-coupling is very hard 
at the LHC. Recently a study of signal and backgrounds has been performed 
for the $g g \to H H$ channel~\cite{slhc}, both for a standard LHC luminosity 
of $10^{34} \, {\rm cm}^{-2} {\rm s}^{-1}$~\cite{bpr02} and for a possible 
future upgrade of the luminosity to $10^{35}\, {\rm cm}^{-2} 
{\rm s}^{-1}$~\cite{slhc}. Among all possible decay channels, the most 
interesting one turned out to be 
$g g \to HH \to W^+ W^- W^+ W^- \to l^\pm \nu j j l^\pm \nu j j$, 
which has a good branching ratio for $m_H \geq 170$~GeV. The like-sign lepton 
requirement is essential to reduce the high-rate opposite-sign lepton final
states from Drell--Yan and $t \bar t$ production. Potential backgrounds 
to the considered signature are given by $t \bar t + $jets, $W Z + $jets, 
$t \bar t W$, $W W W j j$ including the resonant channel $W(H\to WW)jj$ and 
$t \bar t t \bar t$.
\begin{table*}[htb]
\begin{center}
\caption{Expected numbers
 of signal and background events after all cuts for the $ gg\to HH\to
  \ 4W \to  l^+ l'^{+} 4j \nu \nu$ final state, for
  $\int {\cal L}=6000$ fb$^{-1}$~\cite{slhc}.}
\label{tab:ggHH}
\begin{tabular}{cccccccc} 
\hline
$m_H$ & Signal & $ t\bar{t} $ & $ W^{\pm} Z $ & 
 $ W^{\pm}W^+W^- $ & $ t\bar{t} W^{\pm} $ & $ t\bar{t} t\bar{t} $ 
& $S/\sqrt{B}$\\
\hline
170~GeV &   350                &  90              & 60             &
   2400              &  1600            & 30   
& 5.4
                          \\ 
200~GeV & 220               & 90             & 60              &
   1500              & 1600           & 30
& 3.8 \\
\hline                          
\end{tabular}
\end{center}
\end{table*}
By applying the cuts described in ref.~\cite{slhc}, the number of events 
for signal and backgrounds are summarized in Table~\ref{tab:ggHH} for 
an integrated luminosity of 6000~fb$^{-1}$, where a signal significance 
of 5.3 (3.8) $\sigma$ for $m_H = 170 \, (200)$~GeV can be reached, 
optimistically assuming that the main parameters of the detector performance 
will remain the same as those expected at $10^{34}\, {\rm cm}^{-2} 
{\rm s}^{-1}$. 
This would lead to a determination of the total production cross-section 
with a statistical uncertainty of $\pm 20\% \, (\pm 26\%)$ for $m_H = 170$~GeV 
($200$~GeV), allowing a determination of $\lambda_{HHH}$ with statistical 
errors of 19\% (25\%)~\cite{slhc}. In the case of $300$~fb$^{-1}$ only 
the non-vanishing of the Higgs self-coupling could be established 
at 95\% C.L. for $150$~GeV $< m_H < 200$~GeV~\cite{bpr02}.

\section{Summary}
During the last few years there has been a dramatic improvement in both 
theoretical and experimental studies of several Higgs boson production 
and decay 
channels at the LHC. On the theoretical side, the corrections at 
NLO (in one case even at NNLO) have been 
calculated for the main Higgs production processes and for many irreducible 
backgrounds. Several LO Monte Carlo event generators based on exact matrix
elements have been developed very recently to give prediction for 
multiparton final states, which represent important backgrounds for several 
Higgs signatures. On the experimental side, complete 
simulations, including full detector simulation, have been carried out for all 
production processes, pointing out the relevance of the weak boson fusion 
processes as discovery channels in the low Higgs mass region. Considering all 
channels, a signal significance above $5\sigma$ over the entire mass
spectrum is well established already with only 10~fb$^{-1}$ of integrated 
luminosity per experiment. A strategy has been designed to study, in a
model-independent way, the Higgs couplings to fermions and bosons, which allows 
also, with little theoretical assumption, an indirect determination of the 
total Higgs width. Recently the potential of the LHC in the determination 
of the Higgs self-coupling has been investigated, but only with an integrated 
luminosity of $6000$~fb$^{-1}$, and in the mass range 
$170 \leq m_H \leq 200$~GeV a quantitative study could be performed. 
\vskip 4pt
The author wishes to thank S.~Catani, F.~Gia\-not\-ti, R.~Harlander, K.~Jakobs, 
M.L.~Mangano and M.T.~Zeyrek for many useful discussions. 
M.L.~Mangano is also acknowledged for a careful 
reading of the manuscript.


\begin{thebibliography}{9}
\bibitem{atl99-15} ATLAS Collaboration, Tech. Rep. CERN/LHCC/99-15, CERN, 1999.
\bibitem{jakobs} K.~Jakobs, talk given at ATLAS week, Clermont-Ferrand, 
June 2002.
\bibitem{zeyrek} H.D.~Yildiz, R.~Kinnunen and M.T.~Zeyrek, CMS Note 2001/050; 
N.~Akchurin {\it et al.}, CMS Note 2002/016.
\bibitem{zknrw00} R.~Kinnunen {\it et al.}, Phys.~Rev.~D62 (2000) 013009. 
\bibitem{grazzini} M.~Grazzini, hep-ph/0209302; 
S.~Catani, D.~de~Florian and M.~Grazzini, JHEP 0105 (2001) 025; and 
0201 (2002) 015.
\bibitem{kilgore} W.B.~Kilgore, these proceedings; 
R.V.~Harlander and W.B.~Kilgore, Phys.~Rev.~D64 (2001) 025; Phys.~Rev.~Lett.~88 
(2002) 201801; C.~Anastasiou and K.~Melnikov, hep-ph/0207004.
\bibitem{tthnlo} W.~Beenakker {\it et al.}, Phys. Rev. Lett. 87 (2001) 201805; 
                 S.~Dawson and L.~Reina, Phys. Rev. Lett. 87 (2001) 201804; 
            S.~Dawson, L.~Reina and D.~Wackeroth, Phys. Rev. D65 (2002) 053017.
\bibitem{diphox} T.~Binoth {\it et al.}, Eur. Phys. J. C16 (2000) 311; 
hep-ph/0203064.
\bibitem{gluglugg} Z.~Bern, L.~Dixon and C.~Schmidt, hep-ph/0206194.
\bibitem{evc} R.K.~Ellis and S.~Veseli, Phys.~Rev.~D60 (1999) 011501; 
J.~Campbell, hep-ph/0105226.
\bibitem{ce} J.~Campbell and R.K.~Ellis, hep-ph/0202176.
\bibitem{fw} S.~Frixione and B.R.~Webber, hep-ph/0204244. 
\bibitem{qcdhjj} V.~Del~Duca {\it et al.}, Nucl.~Phys.~B616 (2001) 367.
\bibitem{alpgen} M.L.~Mangano {\it et al.}, hep-ph/0206293.
\bibitem{comphep} A.~Pukhov {\it et al.}, hep-ph/9908288.
\bibitem{grace} T.~Ishikawa {\it et al.}, MINAMI-TATEYA Group Coll., KEK-92-19; 
S.~Tsuno {\it et al.}, hep-ph/0204222.
\bibitem{madgraph} T.~Stelzer and W.F.~Long, Comput.~Phys.~Commun.~81 (1994) 357; 
F.~Maltoni and T.~Stelzer, hep-ph/0208156.
\bibitem{acermc} B.P.~Kersevan and E.~Richter-Was, hep-ph/0201302.
\bibitem{cms94-38} CMS Collaboration, Tech. Rep. CERN/LHCC/94-38, CERN, 1994.
\bibitem{dreiner} M. Dittmar, and H.K. Dreiner, Phys.~Rev.~D55 (1997) 167; 
hep-ph/9703401.
\bibitem{denegri} D.~Denegri {\it et al.}, hep-ph/0112045.
\bibitem{rz97} D.~Rainwater and D.~Zeppenfeld,  JHEP 12 (1997) 005.
\bibitem{rzh99} D.~Rainwater, D.~Zeppenfeld and K.~Hagiwara, Phys. Rev. D59
(1999) 014037.
\bibitem{prz00} T.~Plehn, D.~Rainwater and D.~Zeppenfeld, Phys. Rev. D61 (2000) 
093005.
\bibitem{rz99} D.~Rainwater and D.~Zeppenfeld, Phys. Rev. D60 (1999) 113004.
\bibitem{kprz01} N.~Kauer {\it et al.}, Phys. Lett. B503 (2001) 113.
\bibitem{rws99} E.~Richter-Was and M.~Sapinski, Acta Phys. Polon. B30 (1999)
1001. 
\bibitem{dmd01} V.~Drollinger, T.~M\"uller and D.~Denegri, hep-ph/0111312.
\bibitem{mrw02} F.~Maltoni, D.~Rainwater and S.~Willenbrock, hep-ph/0202205.
\bibitem{rb02} A.~Belyaev and L.~Reina, hep-ph/0205270.
\bibitem{dmd02} V.~Drollinger, T.~M\"uller and D.~Denegri, hep-ph/0201249.
\bibitem{z02} D.~Zeppenfeld, hep-ph/0203123.
\bibitem{dkmz99} A.~Djouadi {\it et al.},  
Eur.~Phys.~J.~C10 (1999) 45.
\bibitem{slhc} F.~Gianotti, M.L.~Mangano, T.~Virdee (conveners), 
hep-ph/0204087; A.~Clark, A.~Blondel and F.~Mazzucato, ATL-PHYS-2002-005.
\bibitem{bpr02} U.~Baur, T.~Plehn and D.~Rainwater, hep-ph/0206024.
\end{thebibliography}
\end{document}